# A functional derivative useful for the linearization of inequality indexes in the design-based framework

**Lucio Barabesi[1], Giancarlo Diana[2] and Pier Francesco Perri[3]**

[1]Department of Economics and Statistics, University of Siena, Piazza San Francesco 7, 53100, Siena (Italy)
[2]Department of Statistical Sciences, University of Padova, Via Cesare Battisti 241, 35121, Padova (Italy)
[3]Department of Economics, Statistics and Finance, University of Calabria,
Via P. Bucci, 87036, Arcavacata di Rende (Italy)

**Abstract.** Linearization methods are customarily adopted in sampling surveys to obtain approximated variance formulae for estimators of nonlinear functions of finite-population totals - such as ratios, correlation coefficients or measures of income inequality - which can be usually rephrased in terms of statistical functionals. In the present paper, by considering the Deville's (1999) approach stemming on the concept of design-based influence curve, we provide a general result for linearizing large families of inequality indexes. As an example, the achievement is applied to the Gini , the Amato, the Zenga and the Atkinson indexes, respectively.

**1. Introduction.** Under the usual design-based approach, let $U$ be a fixed population of identifiable individuals labeled (at least ideally) by the first $N$ integers, *i.e.* $U = \{1, \ldots, N\}$, and let $y_i$ be the variable value on the $i$-th individual. In this setting, Deville (1999) has considered the discrete measure on $\mathbb{R}$

$$M = \sum_{i \in U} \delta_{y_i} ,$$

where $\delta_y$ represents the Dirac mass at $y$. Deville (1999) has emphasized that the target population parameter may be generally written as a functional $F$ with respect to $M$, namely $F(M)$. In this case, by supposing that a sample $S$ of size $n$ is selected from $U$ according to a design with first-order and second-order inclusion probabilities respectively given by $\pi_i$ and $\pi_{ij}$, the empirical measure corresponding to $M$ is given by

$$\widehat{M} = \sum_{i \in S} \frac{1}{\pi_i} \delta_{y_i}$$

and the substitution estimator for $F(M)$ may be obtained as $F(\widehat{M})$. If $F$ is homogeneous of degree $\alpha$, under broad assumptions Deville (1999) has proven the linearization

$$\sqrt{n} N^{-\alpha}(F(\widehat{M}) - F(M)) = \sqrt{n} N^{-\alpha} \int \mathrm{IF}_F(u; M) \mathrm{d}(\widehat{M} - M)(u) + o_p(1) ,$$

where

$$\mathrm{IF}_F(u; M) = \lim_{t \to 0} \frac{1}{t} \left( F(M + t\delta_u) - F(M) \right)$$

is the influence function in the design-based approach (see also Goga *et al.*, 2009). From a mathematical perspective, $\mathrm{IF}_F(u; M)$ is actually the Gâteaux differential of $F(M)$ in the direction of the Dirac mass at $u$. Hence, the role of $\text{IF}_F(u; M)$ is central, expecially with the aim of variance estimation for the empirical functional $F(\widehat{M})$ (Deville, 1999).

In this setting, let us consider the functional which may be expressed as

$$F(M) = \int \psi_y(L_y(M)) \, \mathrm{d}M(y) \,, \tag{1}$$

where $L_y(M) = (L_{1,y}(M), \ldots, L_{k,y}(M))^\mathrm{T}$ is a vector of functionals (eventually) indexed by $y$ and $\psi_y : \mathbb{R}^k \mapsto \mathbb{R}$ is a function family assumed to be differentiable and regularly indexed by $y$. The inequality measures commonly considered in practice are members of the functional family $F$, or may be expressed at most as $\varphi(F(M)) = (\varphi \circ F)(M)$ where $\varphi : \mathbb{R} \mapsto \mathbb{R}$ is a smooth function - in the next Section some illustrative examples are provided.

In order to obtain the linearization of the functional $F$ defined in expression (1), the following results are useful. Lemma 1 is introduced for the sake of completeness, since its proof (and the connected assumptions) are not usually reported in statistical literature in its precise version. For more details on Gâteaux and Fréchet differentiability, see *e.g.* Behmardi and Nayeri (2008) and the references therein.

**Lemma 1.** *If $L(M) = (L_1(M), \ldots, L_k(M))^\mathrm{T}$ is a vector of functionals and $\phi : \mathbb{R}^k \mapsto \mathbb{R}$ is a differentiable function, let us consider the functional $\phi(L(M)) = (\phi \circ L)(M)$. By assuming that $L_j$ is Fréchet differentiable for each $j$, the influence function of $(\phi \circ L)$ is given by*

$$\text{IF}_{\phi \circ L}(u; M) = \nabla \phi(L(M))^\mathrm{T} \text{IF}_L(u; M) \,,$$

*where $\text{IF}_L(u; M) = (\text{IF}_{L_1}(u; M), \ldots, \text{IF}_{L_k}(u; M))^\mathrm{T}$.*
**Proof.** By Peano's form of Taylor's formula it holds

$$\phi(L(M + t\delta_u)) = \phi(L(M)) + \nabla \phi(L(M))^\mathrm{T}(L(M + t\delta_u) - L(M)) \\ + o(\|L(M + t\delta_u) - L(M)\|)$$

and, since $o(\|L(M + t\delta_u) - L(M)\|) = o(t)$, by definition it follows that

$$\begin{aligned}\text{IF}_{\phi \circ L}(u; M) &= \lim_{t \to 0} \frac{1}{t} \left( \phi(L(M + t\delta_u)) - \phi(L(M)) \right) \\ &= \lim_{t \to 0} \frac{1}{t} \left( \nabla \phi(L(M))^\mathrm{T}(L(M + t\delta_u) - L(M)) + o(t) \right) \\ &= \nabla \phi(L(M))^\mathrm{T} \lim_{t \to 0} \frac{1}{t} \left( L(M + t\delta_u) - L(M) \right) = \nabla \phi(L(M))^\mathrm{T} \text{IF}_L(u; M)\end{aligned}$$

since $L$ is assumed to be Fréchet differentiable. □

**Proposition 1.** *Let $F$ be the functional defined in (1). If $L_{j,y}$ is Fréchet differentiable for each $j$, the influence function of $F$ is given by*

$$\text{IF}_F(u; M) = \psi_u(L_u(M)) + \int \nabla \psi_y(L_y(M))^\mathrm{T} \text{IF}_{L_y}(u; M) \, \mathrm{d}M(y) \,,$$

*where $\text{IF}_{L_y}(u; M) = (\text{IF}_{L_{1,y}}(u; M), \ldots, \text{IF}_{L_{k,y}}(u; M))^\mathrm{T}$.*
**Proof.** By definition it holds

$$\mathrm{IF}_F(u;M) = \lim_{t\to 0}\frac{1}{t}\left(F(M+t\delta_u) - F(M)\right)$$
$$= \lim_{t\to 0}\frac{1}{t}\left(\int \psi_y(L_y(M+t\delta_u))\,\mathrm{d}(M+t\delta_u)(y) - \int \psi_y(L_y(M))\,\mathrm{d}M(y)\right)$$
$$= \lim_{t\to 0}\frac{1}{t}\int \left(\psi_y(L_y(M+t\delta_u)) - \psi_y(L_y(M))\right)\mathrm{d}M(y) + \lim_{t\to 0}\psi_u(L_u(M+t\delta_u)).$$

For a fixed $u$, we have
$$\lim_{t\to 0}\psi_u(L_u(M+t\delta_u)) = \psi_u(L_u(M)),$$

and, since $\psi_y$ is continuously indexed by $y$, it reads
$$\mathrm{IF}_F(u;M) = \int \lim_{t\to 0}\frac{1}{t}\left(\psi_y(L_y(M+t\delta_u)) - \psi_y(L_y(M))\right)\mathrm{d}M(y) + \psi_u(L_u(M))$$
$$= \int \nabla\psi_y(L_y(M))^{\mathrm{T}}\mathrm{IF}_{L_y}(u;M)\,\mathrm{d}M(y) + \psi_u(L_u(M)),$$

on the basis of the differentiability of $\psi_y$ and Lemma 1. $\square$

Hence, Proposition 1 provides a simple rule for obtaining the influence function corresponding to the functional (1). Finally, it should be remarked that, by means of Lemma 1, the influence function of $(\varphi \circ F)$ also follows, *i.e.*
$$\mathrm{IF}_{\varphi\circ F}(u;M) = \varphi'(F)\mathrm{IF}_F(u;M),$$

by assuming that $\varphi$ be differentiable.

**2. Application to some inequality indexes.** We show the usefulness of Proposition 1 for the linearization of some inequality measures commonly adopted in practice (for a general treatment of inequality indexes, see the classical monograph by Cowell, 2011). As a first example, the celebrated concentration index is considered. In this setting, Langel and Tillé (2012a) have provided the expression of the influence function by developing an *ad hoc* differential rule. The same result may be simply achieved by using Proposition 1. Indeed, the finite-population version of the concentration index (see *e.g.* Berger, 2008) may be expressed as the functional
$$G(M) = \int \frac{2yH_y(M)}{N(M)T(M)}\,\mathrm{d}M(y) - 1,$$

where
$$N(M) = \int \mathrm{d}M(x)$$

is actually the population size $N$ rephrased as a functional and
$$T(M) = \int x\,\mathrm{d}M(x)$$

is the population total. In addition, by assuming that $I_B$ is the usual indicator function of a set $B$,

$$H_y(M) = \int I_{[x,\infty[}(y) \, \mathrm{d}M(x)$$

actually represents the number of individuals whose variable value is less than or equal to a given $y$. In the following, we also assume that

$$K_y(M) = \int x I_{[y,\infty[}(x) \, \mathrm{d}M(x),$$

which obviously turns out to be the total of the variable values greater than or equal to a given $y$. Hence, in this case we have $L_y(M) = (H_y(M), N(M), T(M))^{\mathrm{T}}$, while

$$\psi_y(L_y(M)) = \frac{2y H_y(M)}{N(M) T(M)}$$

and $\varphi(F) = F - 1$. Thus, with a slight abuse in notation, *i.e.* by suppressing the argument of the functionals for the sake of simplicity, it holds that

$$\nabla \psi_y(L_y(M)) = \frac{2y}{NT} \left(1, -\frac{H_y}{N}, -\frac{H_y}{T}\right)^{\mathrm{T}}$$

and $\mathrm{IF}_{L_y}(u; M) = (I_{[u,\infty[}(y), 1, u)^{\mathrm{T}}$. Hence, by applying Proposition 1, after some algebra it follows that

$$\mathrm{IF}_G(u; M) = \frac{2}{NT}(u H_u + K_u) - (G+1)\left(\frac{1}{N} + \frac{u}{T}\right),$$

which coincides with the expression given by Langel and Tillé (2012a).

As a second example, we consider the Amato index, which has recently received renewed interest for its properties (Arnold, 2012). The influence function for the Amato index is not available in literature. To this aim, on the basis of the continuous-population expression of the Amato index (Arnold, 2012), the finite-population counterpart of this inequality measure may be given as the functional

$$A(M) = \int \sqrt{\frac{1}{N(M)^2} + \frac{y^2}{T(M)^2}} \, \mathrm{d}M(y).$$

Hence, in this case $L_y(M) = (N(M), T(M))^{\mathrm{T}}$, while

$$\psi_y(L_y(M)) = \sqrt{\frac{1}{N(M)^2} + \frac{y^2}{T(M)^2}}$$

and, trivially, $\varphi(F) = F$. By adopting the same notational simplification as above and by assuming that $\mu = T/N$ is the population mean, it holds that

$$\nabla \psi_y(L_y(M)) = \frac{T}{\sqrt{\mu^2 + y^2}} \left(-\frac{1}{N^3}, -\frac{y^2}{T^3}\right)^{\mathrm{T}}$$

and $\mathrm{IF}_{L_y}(u; M) = (1, u)^{\mathrm{T}}$. Hence, by applying Proposition 1, it turns out that

$$\mathrm{IF}_A(u;M) = \frac{1}{T}\sqrt{\mu^2+u^2} - \frac{\mu}{N^2}\int \frac{1}{\sqrt{\mu^2+y^2}}\,\mathrm{d}M(y) - \frac{u}{T^2}\int \frac{y^2}{\sqrt{\mu^2+y^2}}\,\mathrm{d}M(y)\,.$$

As a third example, we consider an inequality measure recently proposed by Zenga (2007) - the so-called Zenga new index - which has received considerable attention (see *e.g.* Langel and Tillé, 2012b). Langel and Tillé (2012b) have introduced the finite-population version of the Zenga new index based on smoothed quantiles and have provided the linearization of the corresponding functional. However, by considering the continuous-population expression as proposed by Zenga (2007, expression (5.6) in his paper), the "natural" finite-population counterpart of the Zenga new index may be given as the functional

$$Z(M) = 1 - \int \frac{(N(M) - H_y(M))(T(M) - K_y(M))}{N(M)H_y(M)K_y(M)}\,\mathrm{d}M(y)\,.$$

Hence, in this case $L_y(M) = (H_y(M), K_y(M), N(M), T(M))^{\mathrm{T}}$, while

$$\psi_y(L_y(M)) = \frac{(N(M) - H_y(M))(T(M) - K_y(M))}{N(M)H_y(M)K_y(M)}$$

and $\varphi(F) = 1 - F$. By adopting in turn the same notational simplification for the argument of the functionals, it holds that

$$\nabla \psi_y(L_y(M)) = \left(-\frac{T-K_y}{H_y^2 K_y},\ -\frac{\mu(N-H_y)}{H_y K_y^2},\ \frac{T-K_y}{N^2 K_y},\ \frac{N-H_y}{NH_y K_y}\right)^{\mathrm{T}}$$

and $\mathrm{IF}_{L_y}(u;M) = (I_{[u,\infty[}(y), uI_{[u,\infty[}(y), 1, u)^{\mathrm{T}}$, from which it follows that

$$\begin{aligned}\mathrm{IF}_Z(u;M) = &-\frac{(N-H_u)(T-K_u)}{NH_u K_u} + \int \frac{I_{[u,\infty[}(y)(T-K_y)}{H_y^2 K_y}\,\mathrm{d}M(y) \\ &+ \mu u \int \frac{I_{[u,\infty[}(y)(N-H_y)}{H_y K_y^2}\,\mathrm{d}M(y) - \frac{1}{N^2}\int \frac{T-K_y}{K_y}\,\mathrm{d}M(y) \\ &- \frac{u}{N}\int \frac{N-H_y}{H_y K_y}\,\mathrm{d}M(y)\,.\end{aligned}$$

For the final example, the Atkinson index is assumed. The finite-population counterpart of this inequality measure may be expressed as the functional

$$A_\epsilon(M) = 1 - \left(\int \frac{y^{1-\epsilon}}{N(M)^\epsilon T(M)^{1-\epsilon}}\,\mathrm{d}M(y)\right)^{1/(1-\epsilon)},$$

where $\epsilon \in [0,1[$. Hence, in this case $L_y(M) = (N(M), T(M))^t$, while

$$\psi_y(L_y(M)) = \frac{y^{1-\epsilon}}{N(M)^\epsilon T(M)^{1-\epsilon}}$$

and $\varphi(F) = 1 - F^{1/(1-\epsilon)}$. Hence, with the usual notational simplification for the argument of the functionals, it holds that

$$\nabla \psi_y(L_y(M)) = \frac{y^{1-\epsilon}}{N^\epsilon T^{1-\epsilon}} \left( -\frac{\epsilon}{N}, -\frac{1-\epsilon}{T} \right)^{\mathrm{T}}$$

and $\mathrm{IF}_{L_y}(u; M) = (1, u)^{\mathrm{T}}$. By applying Proposition 1 and after some algebra, it follows

$$\mathrm{IF}_{A_\epsilon}(u; M) = \frac{1 - A_\epsilon}{N} \left( -\frac{u^{1-\epsilon}}{(1-\epsilon)\mu_{1-\epsilon}} + \frac{u}{\mu} + \frac{\epsilon}{1-\epsilon} \right) ,$$

where $\mu_r = N^{-1} \int y^r \, \mathrm{d}M(y)$ represents the $r$-th population moment.